\documentclass[twocolumn, showpacs, longbibliography, prl,aps, amsmath, amssymb,superscriptaddress]{revtex4-1}
\usepackage{amsfonts} 
\usepackage{graphicx} 
\usepackage[colorlinks,linkcolor=blue,citecolor=blue,urlcolor=blue]{hyperref}
\usepackage{float}
\usepackage[normalem]{ulem}
\usepackage[switch,columnwise]{lineno}

\begin{document}

\title{\Large{Universal power law decay in the dynamic hysteresis of an optical cavity with non-instantaneous photon-photon interactions}}

\author{Z. Geng}
\affiliation {Center for Nanophotonics, AMOLF, Science Park 104, 1098 XG Amsterdam, The Netherlands}

\author{K. J. H. Peters}
\affiliation {Center for Nanophotonics, AMOLF, Science Park 104, 1098 XG Amsterdam, The Netherlands}

\author{A. A. P. Trichet}
\affiliation {Department of Materials, University of Oxford, Parks Road, Oxford OX1 3PH, UK}

\author{K. Malmir}
\affiliation {Department of Materials, University of Oxford, Parks Road, Oxford OX1 3PH, UK}

\author{R. Kolkowski}
\affiliation {Center for Nanophotonics, AMOLF, Science Park 104, 1098 XG Amsterdam, The Netherlands}

\author{J. M. Smith}
\affiliation {Department of Materials, University of Oxford, Parks Road, Oxford OX1 3PH, UK}

\author{S. R. K. Rodriguez} \email{s.rodriguez@amolf.nl}
\affiliation {Center for Nanophotonics, AMOLF, Science Park 104, 1098 XG Amsterdam, The Netherlands}

\date{\today}

\begin{abstract}

We investigate, experimentally and theoretically, the dynamic optical hysteresis of a coherently driven cavity with non-instantaneous photon-photon interactions. By scanning the frequency detuning between the driving laser and the cavity resonance at different speeds across an optical bistability, we find a hysteresis area that is a non-monotonic function of the scanning speed. As the scanning speed increases and approaches the memory time of the photon-photon interactions, the hysteresis area decays following a power law with exponent -1. The exponent of this power law is independent of the system parameters. To reveal this universal scaling behavior theoretically, we  introduce a memory kernel for the interaction term in the standard driven-dissipative Kerr model. Our results offer new perspectives for exploring non-Markovian dynamics of light using  arrays of bistable cavities with low quality factors, driven by low laser powers, and at room temperature.

\end{abstract}

\maketitle

Photons in a nonlinear cavity can undergo phase transitions akin to condensed matter systems. Since the seminal works by Graham \& Haken~\cite{Graham70}, Roy \& Mandel~\cite{Roy80OC}, and Scully~\cite{Scully99}, lasers have  inspired numerous studies of phase transitions of light.  Recently, coherently driven cavities supporting mean-field bistability --- two steady-states at a single driving condition --- have taken a central role in studies of photonic phase transitions~\cite{LeBoite, Carmichael15,  Mendoza16, Wilson16, Fitzpatrick17, Fink17PRX, Foss17, Casteels17, Biondi17, Rodriguez17, Angerer17, Fink18, Vicentini18}. Progress in this field has been recently accelerated by three developments. First, various highly nonlinear photonic resonators, and novel methods to probe their dynamics, are  becoming available~\cite{Fitzpatrick17, Fink17PRX, Rodriguez17, Angerer17, Fink18}. Second, fresh insights coupled to novel theoretical methods have revealed that nonlinear cavities can be driven into intriguing non-equilibrium phases~\cite{LeBoite, Finazzi15, Sieberer16, Hartmann16, Noh16, Wilson16, Biondi17, Foss17}.  Third, there is a growing interest in performing combinatorial optimization~\cite{Leleu17, Leleu19, Liew19} and neuromorphic computing~\cite{Opala19} with bistable cavity arrays.

Descriptions of  bistable optical cavities commonly assume instantaneous photon-photon interactions. In the mean-field equation of motion for the intracavity field $\alpha$, this assumption manifests as a Kerr nonlinearity of the form $|\alpha|^2 \alpha$~\cite{Drummond80}. The same cubic nonlinearity is found in the Gross-Pitaevskii equation employed in atomic physics~\cite{Gross61, Pitaevskii61, Dalfovo99}, in the Ginsburg-Landau theory of superconductivity~\cite{GL50}, in the Lugiato-Lefever equation describing pattern formation in nonlinear optics~\cite{LL87}, and in the force derived from Goldstone's Mexican hat potential $V= -|\phi|^2  + |\phi|^4 $ for the scalar field $\phi$ at the heart of the Higgs mechanism~\cite{Goldstone}. In optics, some of the strongest Kerr nonlinearities arise in semiconductor cavities where exciton-exciton interactions are effectively instantaneous~\cite{CarusottoRMP}. A drawback of those cavities is that optical bistability based on Kerr nonlinearities is typically only observed at cryogenic temperatures. In contrast, several optical resonators with relatively slow but strong thermal nonlinearities have routinely displayed bistability at room temperature ~\cite{Lipson04, Carmon04, Notomi05, Priem05, Shi14, Brunstein09, Sodagar15}.  As bona fide bistable systems, thermo-optical resonators may open up new perspectives for  classical Hamiltonian simulation and computation~\cite{Leleu17, Berloff18, Leleu19, Liew19, Opala19} at room-temperature. However, the influence of the thermal relaxation time on the dynamic hysteresis of  bistable cavities remains to be addressed.

\begin{figure}[!]
 \centerline{{\includegraphics[width=\linewidth]{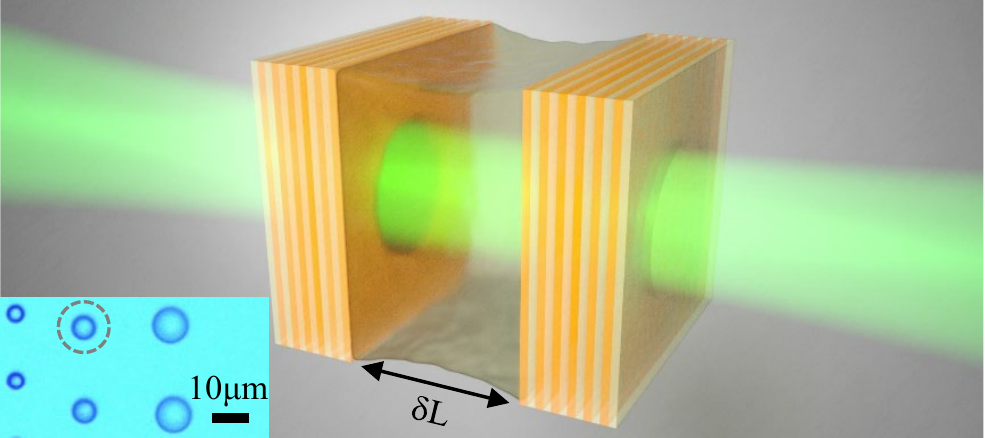}}}\caption{Schematic of a planar-concave micro-cavity filled with oil, as in our experiments. The cavity is illuminated by a continuous wave laser. Inset: Optical image of a chip containing concave mirrors of different size. We used the mirror enclosed by the dashed circle for all experiments. }
\label{fig1}
\end{figure}

In this Letter, we demonstrate signatures of non-instantaneous photon-photon interactions in  the dynamic hysteresis  of a tunable micro-cavity. We investigate a laser-driven micro-cavity filled with oil as shown Fig.~\ref{fig1}, operating at room-temperature. This cavity supports optical bistability at low driving powers $P \sim 70$  $\mu$W. Scanning the cavity length under laser illumination, we observe an optical hysteresis that depends pronouncedly on the ratio of the scanning time to the memory time of the interactions. In contrast to previous reports of dynamic hysteresis in resonators with effectively instantaneous interactions~\cite{Mandel90, Casteels16, Rodriguez17, Lagoudakis18}, we find a hysteresis area that is a non-monotonic function of the scanning speed. For fast scans, the hysteresis area decays following a power law with a universal exponent independent of the system parameters. Our results elucidate how the hysteretic behavior characterizing first-order phase transitions, and the boundary between phases, dynamically vanish when the nonlinearity has a finite memory time.

Figure~\ref{fig1} illustrates our system: a tunable Fabry-P\'{e}rot micro-cavity  driven by a 532 nm continuous wave laser. The cavity is made by a concave and planar mirror, each comprising a distributed Bragg reflector (DBR) on a glass substrate. The mirrors have a peak reflectance of $99.9\%$ at 530 nm, which is the center of the  stop-band. The concave mirror was fabricated by milling a glass substrate with a focused-ion beam prior to the deposition of the DBR~\cite{Trichet15}. The Fig.~\ref{fig1} inset shows a chip containing concave mirrors suitable for making micro-cavities with different mode volumes.  In this work, we use the mirror enclosed by the dashed circle in Fig.~\ref{fig1}, which has a diameter of  $7$  $\mu$m and a radius of curvature of $12$ $\mu$m. Thanks to the strong lateral confinement and high mirror reflectivity we can probe single optical modes across cavity length scans of several nanometers.

The chip containing the concave mirror is aligned parallel to the planar mirror using a hexapod nanopositioner.  This nanopositioner controls all three translational (rotational) degrees of freedom of the concave mirror with nanometer (micro-degree) precision. The planar mirror is mounted on another actuator used to scan the cavity length. Optical excitation and collection are achieved through $10\times$ microscope objectives with numerical aperture $NA=0.25$. The cavity transmission is measured by a photodetector and an oscilloscope. Further details about our setup are included in supplemental information~\cite{supp}.

\begin{figure}[!]
 \centerline{{\includegraphics[width=\linewidth]{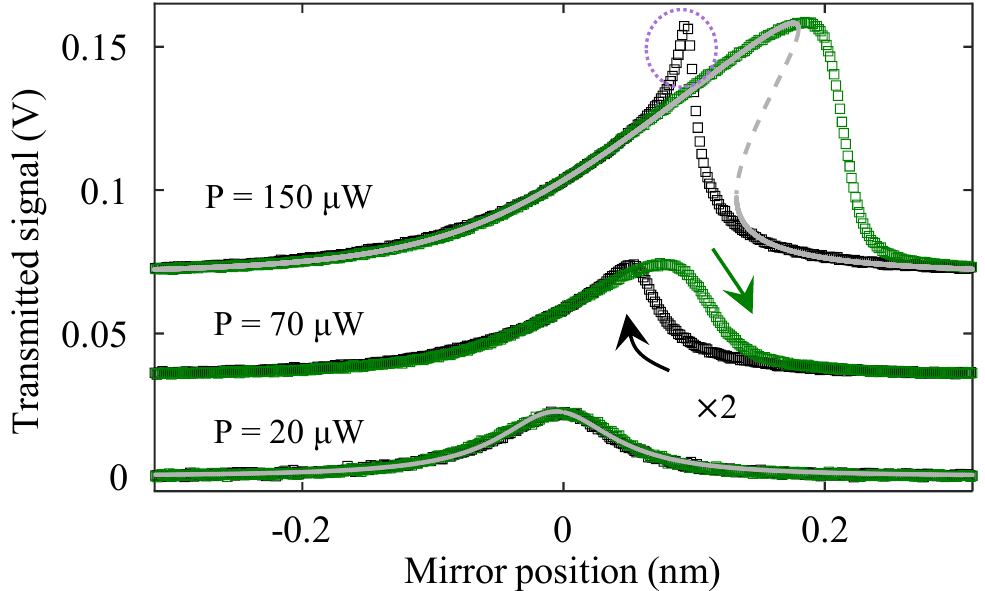}}}\caption{ Average dynamic hysteresis measured by scanning $\Delta/\Gamma$ (see Eq.~\ref{eq1}) at constant speed for three driving powers $P$. The green (black) curve corresponds to opening (closing) the cavity. The overshoot enclosed by the dotted circle emerges from non-instantaneous photon-photon interactions. For clarity, we multiplied the measurements for $P=20$ $\mu$W by 2 and vertically displaced the other measurements. Gray curves are calculations with Eq.~\ref{eq1} as explained in the text.}
\label{fig2}
\end{figure}

To endow the cavity  with a nonlinear optical response, we placed a drop of olive oil inside. Oils are known for their thermo-optical nonlinearities~\cite{Dreischuh, Souza, Garcia15}.  Through z-scan measurements we estimated the nonlinear refractive index $n_2$ of our olive oil to be $\sim -5 \times 10^{-8}$ cm$^2$/W at $532$ nm, consistent with Ref.~\cite{Garcia15}.  Figure~\ref{fig2} shows the transmitted intensity through our oil-filled cavity averaged over 70 cycles and at three laser powers.  Green and black data points  correspond to opening and closing the cavity, respectively. For low powers $P \lesssim 20$  $\mu$W, the cavity response is linear.
The gray curve over the measurements for $P =20$  $\mu$W is a Lorentzian fit, yielding a resonance linewidth of $0.104 \pm 0.001$ nm. For  $P = 70$  $\mu$W, the transmission displays hysteresis (see arrows in Fig.~\ref{fig2}) and a narrow bistability around a mirror position of 0.1 nm. The power needed for bistability in our cavity is similar to that in state-of-the-art monolithic semiconductor cavities~\cite{Deveaud15, Rodriguez17, Fink18}, but at conveniently lower quality factors ( by a factor of $\sim10$) and operating at room-temperature instead of $\sim 5$ K.  For $P = 150$  $\mu$W, the bistability and hysteresis range enlarge as expected. An estimate of  the maximum temperature rise in our bistable oil-filled cavity is provided in supplemental information~\cite{supp}.

All measurements in  Fig.~\ref{fig2} correspond to linear ramps of the cavity length at $1.75$ $\mu$m/s.  Already for this relatively slow scan, a pronounced overshoot followed by a slow decay of the transmitted intensity arises when closing the cavity in the nonlinear regime. This overshoot is due to the finite thermo-optical response time of the cavity, which is not captured by the standard driven-dissipative Kerr model for a single-mode cavity with instantaneous interactions~\cite{Carmon04}.

The standard Kerr model for the intra-cavity mean-field $\alpha$ in a frame rotating at the driving frequency $\omega$ is

\begin{equation}\label{eq1}
i\dot{\alpha}=\left(-\Delta - i \frac{\Gamma}{2} + U (\vert \alpha \vert^2 -1) \right)\alpha + i \sqrt{\kappa_{1}}F.
\end{equation}

\noindent $\Delta=\omega-\omega_0$ is the laser-cavity detuning, with $\omega_0$ the resonance frequency. $U$ is the photon-photon interaction strength. $F$ is the driving amplitude. The total loss rate $\Gamma = \kappa_1 + \kappa_2 + \gamma$ is the sum of the input-output leakage rates through the two mirrors, $\kappa_{1,2}$, and the intrinsic cavity loss rate $\gamma$ due to absorption.  The steady-state follows from setting $\dot{\alpha}=0$ in Eq.~\ref{eq1}.

We attempted to fit the steady-state photon density $|\alpha|^2$  calculated with Eq.~\ref{eq1} to the measurements for $P = 150$  $\mu$W in Fig.~\ref{fig2}, with $F$ as the only relevant adjustable parameter~\cite{supp}. $\Gamma$ is fixed  by the resonance linewidth observed in the linear regime. Furthermore,  since Eq.~\ref{eq1} is a mean-field model, the absolute number of photons $|\alpha|^2$ or the interaction energy $U$ alone are irrelevant; the spectral lineshape  is determined by the ratio  $U|\alpha|^2/\Gamma$. Therefore, any spectral lineshape can be obtained by varying $F$ for any fixed $U$ and $\Gamma$. Consequently, we adjusted $F$ until obtaining the gray curve plotted over the measurements for $P = 150$  $\mu$W. Solid and dashed curves represent stable and unstable states, respectively, determined following Ref.~\onlinecite{Drummond80}.  The fit is good far from resonance, but deviates from the data near the bistability. Next, we show that our data deviates more pronouncedly from predictions of the standard Kerr model as the scanning speed increases.

\begin{figure}[!]
 \centerline{{\includegraphics[width=\linewidth]{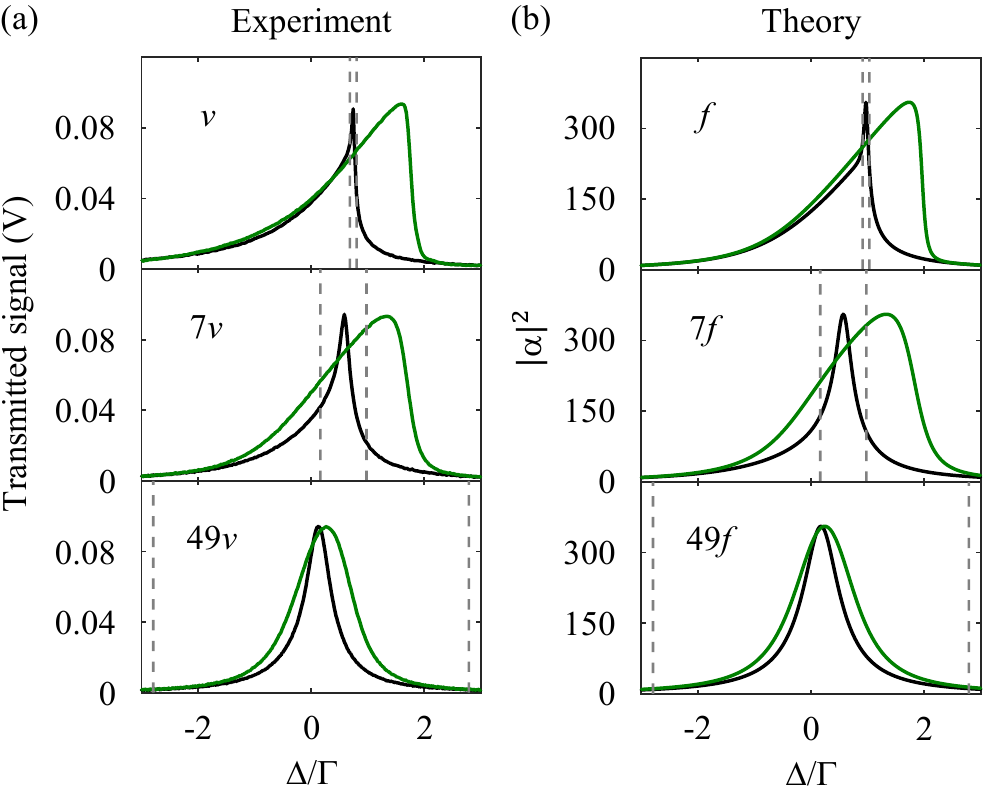}}}\caption{ (a) Measurements and (b) calculations of  average dynamic hysteresis when $\Delta/\Gamma$ is scanned at three different speeds and constant power. The slowest scanning speed is $\nu=0.74$ $\mu$m/s in (a), and $f=5.97 \times 10^{-6}$ $\Gamma^2$ in (b). The power is $P=150$ $\mu$W in (a) and $F=1.52 F_c$, with $F_c$ the critical amplitude  needed for bistability, in (b); note that the power in (a) also corresponds to $F=1.52 F_c$. Measurements in (a) are averaged over $70$ realizations. The vertical dashed lines in (a) and (b) indicate the range of $\Delta/\Gamma$  corresponding to the thermal relaxation time $\tau$ (see text for details).}
\label{fig3}
\end{figure}

We performed hysteresis measurements  for $P = 150$  $\mu$W and various scanning speeds.  We selected a laser power far above the bistability  threshold to limit the influence of noise on our measurements.  Figure~\ref{fig3}(a) shows average dynamic hysteresis measurements for three speeds. Top-to-bottom, the speed is $\nu$, $7\nu$, and $49\nu$, with $\nu=0.74$ $\mu$m/s.  The transmitted intensity is shown as a function of $\Delta/\Gamma$, which we determined from the mirror position and the resonance linewidth in Fig.~\ref{fig2}. Figure~\ref{fig3}(a) shows how the hysteresis cycle qualitatively changes with the scanning speed. Increasing the speed from $\nu$ to $7\nu$ makes the overshoot broader and the hysteresis wider. Interestingly, further increasing the speed to $49\nu$ makes the overshoot broader but the hysteresis narrower.  The measured lineshape for  $49\nu$ resembles a  Lorentzian resonance for both scanning directions, although small deviations exist. This resemblance suggests that the response of the cavity is mostly linear for fast scans, regardless of the high power.

The behavior in Fig.~\ref{fig3}(a) can be explained by considering the finite heating and cooling time of our oil-filled micro-cavity; this makes photon-photon interactions non-instantaneous. Therefore, we  modify Eq.~\ref{eq1} by letting

\begin{equation} \label{subs}
U\left(|\alpha(t)|^2-1\right) \rightarrow \int_0^t\mathrm{d}s\; K(t-s)\left(|\alpha(s)|^2-1\right)\equiv w(t),
\end{equation}

\noindent with the kernel function defined as $K(t)=\frac{U}{\tau}\mathrm{e}^{-t/\tau}$.  $\tau$ is the memory time of the nonlinearity, which corresponds to the thermal relaxation time of our micro-cavity.  Here we have followed the prescription of Mori~\cite{Mori} and H\"anggi~\cite{Hanggi78} for dealing with finite-time interactions. However, whereas Mori-type equations involve non-instantaneous dissipation, we introduced non-instantaneous photon-photon interactions.

Making the substitution~\ref{subs} in Eq.~\ref{eq1} yields an integro-differential equation, which can be conveniently written (for numerical simulation) as two coupled differential equations:

		\begin{subequations}\label{2eqns}
			\begin{eqnarray}
				i \dot{\alpha}(t)=\left(-\Delta- i\frac{\Gamma}{2} + w(t)\right)\alpha(t) + i \sqrt{\kappa_1}F,  \label{eqn2a}
  \\
				\dot{w}(t)=\left[U\left(|\alpha(t)|^2-1\right)-w(t)\right]/\tau.  \label{eqn2b}
			\end{eqnarray}
		\end{subequations}

Equations~\ref{subs} and~\ref{2eqns} imply that the state of the system depends on its entire past, weighted by the memory kernel $K(t)$. Thus, photon-photon interactions are non-local in time. Note that when $\alpha(t)$ is constant and can be taken out of the integral in~\ref{subs}, we recover  Eq.~\ref{eq1}. Hence, steady-states are unchanged by $K(t)$.

Figure~\ref{fig3}(b) shows dynamic hysteresis  calculations using Equations~\ref{2eqns}, with the same parameter values used for the steady-state calculations in Fig.~\ref{fig2}. As for the experiments, we show scanning speeds a factor of 7 apart. The model faithfully reproduces all features observed in experiments.   In the calculations, we set the memory time to $\tau=10^4\Gamma^{-1}$ and the slowest scanning speed to $f=5.97 \times 10^{-6}$ $\Gamma^2$~\cite{supp}.   Relative to the experiments, the value of $\tau$ is smaller (details ahead) and the speed is larger. We rescaled the time scales to avoid unnecessarily long and memory-expensive calculations. Our mean-field calculations can be directly compared to experiments because we respect the hierarchy of time scales in experiments: $\Gamma^{-1} \ll \tau \lesssim T_{b}$, with $T_{b}$ the scanning time across the bistability. Moreover,the ratio $T_{b} / \tau$ is similar for experiments and calculations.

\begin{figure}[!]
 \centerline{{\includegraphics[width=\linewidth]{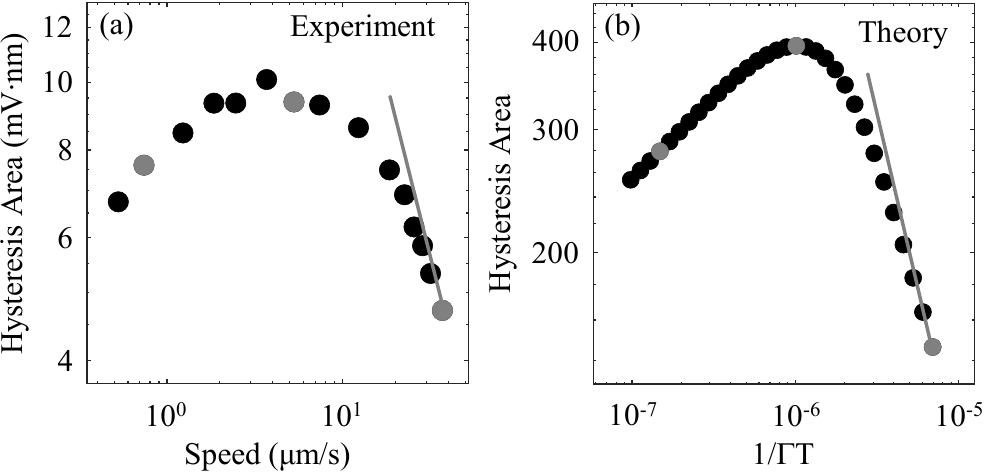}}}\caption{ (a) Measured and (b) calculated hysteresis area as a function of the scanning speed. The dynamic range is the same in (a) and (b). Measurements in (a) are averaged over $70$ realizations. Gray data points correspond to the speeds presented in Fig.~\ref{fig3}. Gray lines are power law fits with exponent -1.}
\label{fig4}
\end{figure}

Next, we analyze the hysteresis area across a range of scanning speeds. The hysteresis area is defined as $A=\int_0^T | I_{\Delta \downarrow}-I_{\Delta \uparrow}  | dt$ , with $I_{\Delta \downarrow }$ and $I_{\Delta \uparrow}$ the transmitted intensity when $\Delta$ decreases and increases, respectively.  $T$ is the driving period, which exceeds $T_b$. In Fig.~\ref{fig4}(a)  we plot the experimental average $A$. The corresponding calculations based on Equations~\ref{2eqns} are presented in Fig.~\ref{fig4}(b). In both measurements and calculations, $A$ peaks at a certain scanning speed.  This peak arises at the cross-over between two dynamical regimes.  For slow scans, $A$ increases with the speed because the cavity cannot adiabatically follow the driving force.  This regime of dynamic hysteresis and the corresponding scaling laws for $A$ have been previously explored experimentally and theoretically~\cite{Mandel90, Roy95, Casteels16, Rodriguez17}. The second and new regime we investigate comprises speeds above the value for which $A$ peaks.  Therein, $A$ decays with increasing speed  because the nonlinearity does not have time to build up during the scan. Essentially, the second regime corresponds to a transition from nonlinear to linear dynamics.  We interpret this transition as an effective reduction of the number of attractors in our system from two to one.

Our measurements are limited to scanning speeds between  $\sim 0.5$ $\mu$m/s and $\sim 40$ $\mu$m/s. The upper speed limit is determined by the resonance frequency of our piezoelectric actuator. On the other end, we limited our measurements to speeds above   $0.5$ $\mu$m/s to avoid low-frequency mechanical noise in our setup. Despite these limitations, the good agreement between experiments and calculations in the available range encourages us to use our model to interpret the physics over an extended speed range.

\begin{figure}[!]
 \centerline{{\includegraphics[width=\linewidth]{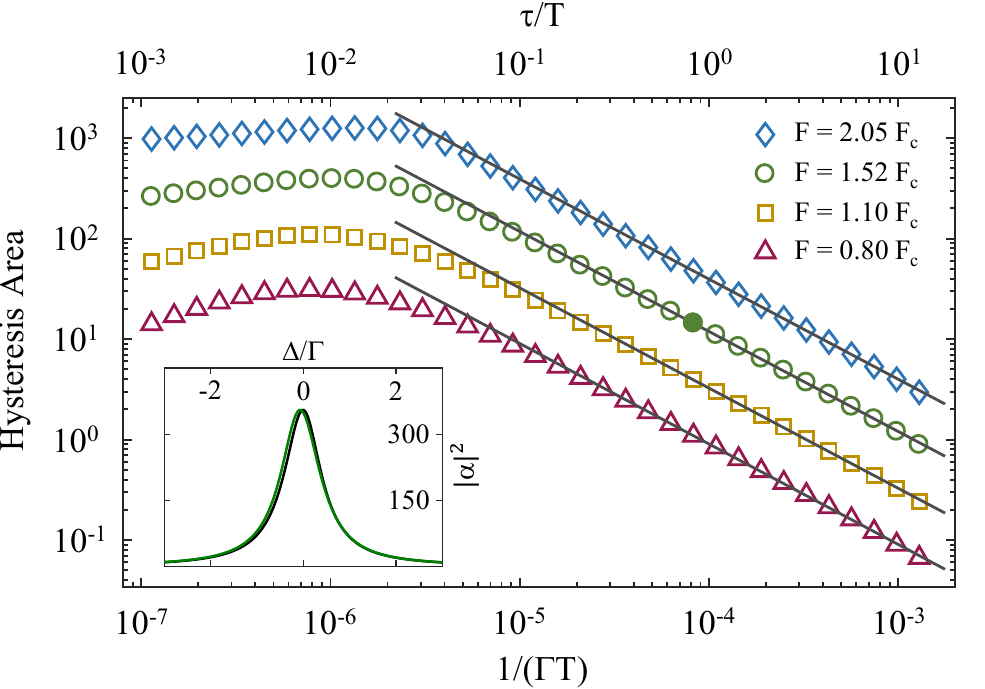}}}\caption{ Calculated hysteresis area as a function of the scanning period $T$. Symbols of different colors correspond to different driving amplitudes $F$ relative to the critical amplitude needed for bistability $F_c$. The bottom and top axis show $T$ referenced to the loss rate $\Gamma$ and the termal relaxation time $\tau$, respectively. Gray lines over the calculations for all $F$ at high speeds are fitted power laws with exponent -1.  Inset: Intracavity photon number $|\alpha|^2$  versus $\Delta/\Gamma$ for the scan indicated by the filled green circle in the main panel, corresponding to $F=1.52 F_c$.}
\label{fig5}
\end{figure}

In Fig.~\ref{fig5} we calculate $A$ across a wide range of speeds for different $F$. At low speeds, the driving conditions determine the scaling of $A$~\cite{Rodriguez17}. At high speeds, we find that $A$ decays following a  power law with universal exponent -1. By `universal' we mean that the exponent, i.e. the slope of the gray lines fitted to the data in Fig.~\ref{fig5},  is independent of the system parameters. To assess whether our experiments show  evidence of this behavior, in  Fig.~\ref{fig4}(a) we plot a power law with exponent -1 over our high-speed data points.  The overlap between this power law and our experimental data for the highest speeds suggests that we reached the onset of the -1 power law regime. For comparison, we plot a -1 power law on top of the corresponding calculations in Fig.~\ref{fig4}(b). In this case, the power law was fitted to the calculations  in Fig.~\ref{fig5} over an extended range. As in experiments, we observe the onset of the -1 power law within the restricted speed range Fig.~\ref{fig4}(b).

Recent calculations~\cite{Casteels16}  and experiments~\cite{Rodriguez17} on dynamic hysteresis in cavities with instantaneous interactions observed a universal power law decay of $A$ at low speeds. In that case, $A$ decays due to the influence of quantum fluctuations. Coincidentally, the universal exponent discovered in Refs.~\cite{Casteels16, Rodriguez17} is also -1, as in the present work.  However, the scaling behavior discovered herein has an entirely different origin (i.e., due to non-instantaneous interactions and unrelated to fluctuations) and arises in the opposite regime of fast scans.

Next, we estimate the experimental thermal relaxation time by comparing experimental and theoretical hysteresis cycles in  Fig.~\ref{fig3}. Since in theory we set $\tau$ and the speed at which $\Delta/\Gamma$ is scanned,  $\tau$  can be converted to a range of $\Delta/\Gamma$ and viceversa.  In Fig.~\ref{fig3}(b) we indicate the range of $\Delta/\Gamma$ corresponding to  $\tau$ by dashed gray lines for the three speeds considered. As expected, the range of $\Delta/\Gamma$ corresponding to $\tau$ increases with the speed.  For the lowest speed $f$,  the scanning time across the bistability range $T_{b}$ largely exceeds $\tau$. In that regime, the overshoot observed when $\Delta$ decreases is the most significant feature which is not captured by the standard Kerr model. The overshoot decays within a time $\sim \tau$, as expected. For the intermediate speed $7f$, $T_{b} \sim \tau$ and the width of the overshoot approaches the bistability range. For the highest speed $49f$, $T_{b} < \tau$ and we have close-to-linear response. Experiments in Fig.~\ref{fig3}(a) display the same behavior as the calculations. Hence, in a similar way we indicate the  range of $\Delta/\Gamma$ corresponding to $\tau$ by two dashed gray lines in Fig.~\ref{fig3}(a). Based on this range of $\Delta/\Gamma$  and our knowledge of the experimental scanning speed, all three measurements in Fig.~\ref{fig3}(a) are consistent with a thermal relaxation time  $\tau = 16 \pm 1$ $\mu$s.

In summary, we have shown how non-instantaneous photon-photon interactions influence the dynamic hysteresis of a coherently driven cavity. Non-instantaneous interactions maximize the hysteresis area at a finite scanning speed. At high speeds, the area decays following a universal power law  with exponent -1. Through this scaling law, the hysteresis characterizing first-order phase transitions vanishes. Beyond single-cavity physics, our observation of optical bistability in oil-filled cavities paves the way for realizing bistable coupled cavities~\cite{Dufferwiel15} and bistable cavity arrays at room-temperature. Such arrays could be used to probe Ising-type phase transitions~\cite{Foss17} and solve combinatorial optimization problems~\cite{Leleu17, Leleu19, Liew19}, or to explore non-Markovian  dynamics in complex optical networks.  Unlike standard non-Markovian systems where finite-time dissipation makes the intracavity noise colored~\cite{Mori, Hanggi78, Hanggi82, Hanggi05, Farias, Farias09}, our system may support non-Markovian dynamics even for white intracavity noise. However, the cavity output noise spectrum is expected to be cut-off at high frequencies by the thermal relaxation time. This new regime of non-Markovian dynamics may be accessed by reducing the laser power so that existing noise in the cavity exerts a greater influence on the bistability, or by injecting tailored noise using modulators~\cite{Deveaud15}.

\section*{Acknowledgments}
This work is part of the research programme of the Netherlands Organisation for Scientific Research (NWO). We thank Ricardo Struik and Niels Commandeur for technical support, and Ewold Verhagen and Femius Koenderink for stimulating discussions. S.R.K.R. acknowledges a NWO Veni grant with file number 016.Veni.189.039.


%

\renewcommand{\thefigure}{S\arabic{figure}}
\setcounter{figure}{0} 
\hyphenation{micro-cavity}

\clearpage
\section{Supplemental Material}
\subsection{Experimental setup}
\begin{figure}[!]
	\centerline{{\includegraphics[width=\linewidth]{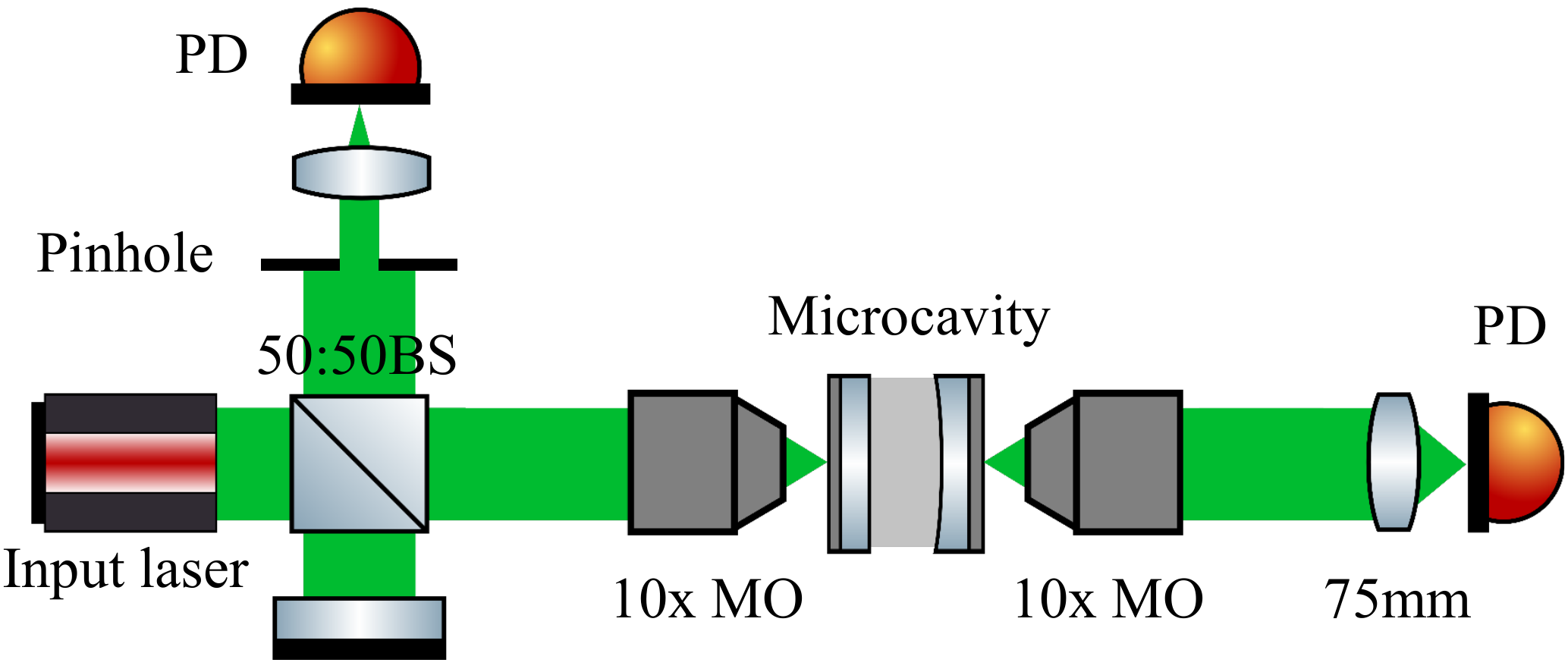}}}\caption{Schematic of the experimental setup. MO is microscope objective, PD is photodetector. }
	\label{fig:optical_setup}
\end{figure}

Figure~\ref{fig:optical_setup}  illustrates the setup we used to optically probe and characterize our oil-filled tunable microcavity. The cavity is driven by a single-mode  continuous wave laser emitting at a wavelength $\lambda = 532$ nm. Excitation and collection are achieved through  $10\times$ microscope objectives with numerical aperture $NA=0.25$. In all our experiments, we drive the fundamental transverse mode of the $9^{th}$ measurable longitudinal mode of the microcavity. Taking into account the electric field penetration into the distributed Bragg reflectors comprising our mirrors, the effective cavity length is $\sim 3$ $\mu$m. To exclude multi-mode interference effects in our measurements, we optimized the in-coupling efficiency of the laser into the desired mode by finely adjusting the position of the concave mirror relative to the laser beam. Finally, the transmitted laser light was focused onto a photodetector by a $f=75$ mm lens.

The cavity length is modulated by displacing one of our mirrors with a  piezoelectric actuator in closed-loop configuration.  Through software, we specify the waveform, frequency, and travel range of the mirror. In all measurements we specified a linear ramp with a travel range of 350 nm.  For  modulation frequencies above  $\sim 50$ Hz, we suspected that the actual travel range of our actuator differed from the specified one. Therefore, we built a  Michelson interferometer in the input arm of the setup [see Fig.\,\ref{fig:optical_setup}] to characterize the mirror displacement. In particular, we measured the time-dependent intensity of a small section of the interferogram while the cavity length was being modulated. Next, we  fitted the  measured total intensity $I_t$ at the output of the interferometer with a two-beam interference equation of the form: $I_t = I_1 + I_2 + 2\sqrt{I_1 I_2}$cos($2kz + \phi$).  $I_1$ and  $I_2$ are the intensities in the two arms of the interferometer, $k=2\pi/\lambda$ is the angular wavenumber, $z$ is the displacement of the actuator in one of the arms, and $\phi$ is a free parameter corresponding to the initial phase difference between the two arms.  Through this analysis, we obtained the actual range traveled by our mirror. Finally, in combination with the scanning time, this travel range was used to calculate the scanning speeds along the horizontal axis of Fig. 4 in the main manuscript.  Once the displacement of the mirror was properly calibrated, the beam splitter used for the Michelson interferometer was removed in order to avoid unwanted reflections which could disturb the hysteresis measurements.

\subsection{Temperature rise in the oil-filled microcavity}
In this section we estimate the temperature rise in our oil-filled microcavity in the nonlinear regime.  To this end, let us first consider the mean-field equation of motion for the coherent field $\alpha$ in a driven-dissipative Kerr nonlinear cavity, i.e. Eq. 1 in the main manuscript. Calculating the steady-state solutions to that equation, one finds that the number of photons in the cavity $N=|\alpha|^2$ satisfies:

\begin{equation}\label{equ:ss_lorentzian_a}
N = \frac{\kappa_1 |F|^2}{\tilde{\Delta}^2 + (\Gamma/2)^2}.
\end{equation}

As in the main manuscript, $\Gamma$ is the total loss rate, $U$ is the photon-photon interaction strength, $\kappa_1$ is the input-output leakage rate, and $F$ is the driving amplitude. In addition, we have defined the quantity

\begin{equation} \label{Dtilde}
\tilde{\Delta}=\Delta - UN,
\end{equation}

\noindent with $\Delta=\omega -\omega_{0}$ the frequency detuning between the driving laser and the cavity resonance, and $UN$ the total interaction energy associated with a population of $N$ photons.

In Eq.~\ref{equ:ss_lorentzian_a}, $N$ appears as a Lorentzian function of the effective detuning $\tilde{\Delta}$. However, since $N$ also enters into the right hand side of Eq.~\ref{equ:ss_lorentzian_a} via $\tilde{\Delta}$, $N$ is a multi-valued function of the driving parameters $F$ and $\Delta$ . Indeed, Eq.~\ref{equ:ss_lorentzian_a} corresponds to a third order polynomial in $N$. This means that, in general, three steady-state solutions exist for a single driving condition. Of these three solutions, at most two are stable; this is known as bistability.  The critical driving amplitude for observing bistability is $F_{c}=\sqrt{\sqrt{3}\Gamma^3/\left(9\kappa_{1}|U|\right)}$.

Next, we analyze the resonant response of our oil-filled microcavity. The cavity resonance frequency is
\begin{equation}
	\omega_{0} = \frac{q c}{2nL}\, ,
\end{equation}
with $c$ the speed of light, $q$ the longitudinal mode number, $n$ the linear refractive index of the intra-cavity medium at ambient temperature, and $L$ the cavity length. When the cavity length is scanned under laser illumination, the resonance frequency changes to
\begin{equation}\label{equ:double_change}
	\tilde{\omega_{0}}=\frac{qc}{2}\frac{1}{(n+\delta n)}\frac{1}{(L+\delta L)}\, ,
\end{equation}
with $\delta n$ the refractive index change due to laser-induced heating of the oil, and $\delta L$ the change in cavity length due to the scan.  Next, we give an approximate expression for  Eq.~\ref{equ:double_change} based on two observations: i) As Fig. 2 in the main article shows,  the hysteresis range spans less than 0.2 nm in mirror displacement even for the highest laser power. Meanwhile,  the initial cavity length is around 3 $\mu$m. Hence, $\delta L \ll L$. ii) $\delta n$ is on the order of $10^{-4}$ \cite{Khodier}. Thus, we have $\delta n \ll n$.  Based on the above two inequalities,
\begin{equation}\label{equ:expand_double_change}
	\tilde{\omega_{0}} \approx \frac{qc}{2nL}\left(1-\frac{\delta n}{n}-\frac{\delta L}{L}\right).
\end{equation}

Next, we insert Eq.~\ref{equ:expand_double_change} into Eq.~\ref{Dtilde}, and we take the value of $L$ that satisfies  $qc/2nL=\omega$. Consequently, we obtain

\begin{equation}\label{equ:detuning_thermal}
	\tilde{\Delta} = \omega\frac{\delta L}{L}+\omega\frac{\delta n}{n}.
\end{equation}

The first term on the right hand side of Eq.~\ref{equ:detuning_thermal} corresponds to the detuning $\Delta$ in Eq.~\ref{Dtilde}. The second term in Eq.~\ref{equ:detuning_thermal} corresponds to the interaction energy, which is proportional to the intensity-induced refractive index change $\delta n$.

For our oil-filled cavity with thermo-optical nonlinearity, $\delta n$ is given by

\begin{equation}\label{equ:thermal_delta}
	\delta n=\frac{dn}{dT}\delta T.
\end{equation}

\noindent with $dn/dT$ a material constant and $\delta T = T-T_{0}$ the temperature change of the oil due to the light intensity.

Next, we estimate $\delta T$ based on energy conservation arguments, similar to Ref.~\onlinecite{Carmon04}. $\delta T$ is related to the heat that goes inside and outside the cavity, i.e. $q_{in}$ and $q_{out}$, via
\begin{equation} \label{cons}
	C \dot{\delta T}= \dot{q}_{in} -  \dot{q}_{out},
\end{equation}

\noindent with $C$ the heat capacity of the oil.  Next, we assume that $\dot{q}_{in}=AN$, with $N$ the number of photons in the cavity and $A$ a constant describing the conversion of absorbed photons into heat. Furthermore, we assume that $\dot{q}_{out}=B  \delta T$, with $B$ a constant describing heat dissipation into the environment.  Hence, Eq.~\ref{cons} becomes

\begin{equation}
	C \dot{\delta T}= A N-B \delta T.
\end{equation}
Therefore, in steady state ($\dot{\delta T}=0$), the temperature rise $\delta T$ is proportional to the photon number $N$:
\begin{equation}\label{equ:SS_T_N}
	\delta T=\frac{A}{B}N.
\end{equation}

Combining Equations~\ref{equ:detuning_thermal}, ~\ref{equ:thermal_delta}, and ~\ref{equ:SS_T_N}, we  find expressions for the rescaled detuning $\tilde{\Delta}$, the linear detuning $\Delta$, and the thermo-optically induced interaction constant $U_{T}$:
\begin{subequations}
	\begin{equation}
	\label{equ:thermal_nonlinear_detuning_final_a}
		\tilde{\Delta}=\Delta -U_{T}N\,,
	\end{equation}
	\begin{equation}
	\label{equ:thermal_nonlinear_detuning_final_b}
		\Delta=\omega \frac{\delta L}{L}\,,
	\end{equation}
	\begin{equation}
	\label{equ:thermal_nonlinear_detuning_final_c}
		U_{T}=-\omega\frac{1}{n}\frac{dn}{dT}\frac{A}{B}.
	\end{equation}
\end{subequations}

Based on the above analysis, we can estimate the temperature rise in the measurement of $P=150$ $\mu$W shown in Fig. 2 of the main article. Equation~\ref{equ:SS_T_N} states that the highest temperature in the measurement corresponds to the largest $N$. From Eq.~\ref{equ:ss_lorentzian_a} the largest $N$ corresponds to $\tilde{\Delta}=0$. Using Eq.~\ref{equ:detuning_thermal} and ~\ref{equ:thermal_delta}, we obtain
\begin{equation}\label{equ:estimation}
	\tilde{\Delta}=\omega\frac{\delta L}{L}+\omega\frac{1}{n}\frac{dn}{dT}\delta T=0
\end{equation}

Finally, we solve the above expression for  $\delta T$, and insert the parameter values corresponding to our experiment with a driving power of 150 $\mu$W. In particular, in Fig. 2 of the main article we observe $\delta L=0.177$ nm. Furthermore, the cavity length is $L \sim 3$ $\mu$m, and the linear refractive index of our oil at $\lambda=532$ nm is $n\approx 1.45$. In addition, based on Ref.~\onlinecite{Khodier} we estimate $dn/dT \sim -4\times10^{-4}$. Inserting these numbers in Eq.~\ref{equ:estimation}, we find the greatest temperature rise is $\delta T=0.2^\circ$ C.

\subsection{Calculation details}
In this section we provide further details about the calculations in the main text, and we explain how parameter values were determined or selected. Let us first consider the steady-state calculations based on Eq. 1 and presented in Fig. 2 of the main text.  In particular, we fitted the steady-state photon number  $|\alpha|^2$  to the experimental transmitted signal ($\propto |\alpha|^2$) for a laser power $P=150$ $\mu$W.  The model parameters are the photon-photon interaction strength $U$,  the total loss rate $\Gamma$,  the driving amplitude $F$, and the input-output leakage rates $\kappa_{1,2}$.  At first sight, it may seem that these five parameters can be freely adjusted in order to fit the measured lineshape. However, as explained next, we do not have this freedom due to several considerations and constraints.

The starting point of our analysis is the realization that  Eq. 1 in the main text is  a mean-field model neglecting quantum fluctuations. Hence, the value of each parameter individually, or of $|\alpha|^2$, is irrelevant.  The spectral lineshape  is entirely determined by the ratio $U |\alpha|^2 / \Gamma$. In particular, the linear regime is characterized by $U |\alpha|^2 \ll \Gamma$. Bistability emerges for $U |\alpha|^2 \gtrsim \Gamma$.

Next, let us explain how the model parameters relevant to the fit in Fig. 2 were set. First, the value of $\Gamma$ was determined by fitting a Lorentzian function to the measured lineshape in the linear regime ($P=20$ $\mu$W). Second, note that $\kappa_1$ is just a multiplicative factor for $F$. In fact, we could have defined an effective driving amplitude $F'= \sqrt{\kappa_1} F$ in Eq. 1 and not introduced $\kappa_1$ at all. We included $\kappa_{1,2}$ in our model for consistency with standard  input-output theory, and to have the right units for $F$. Therefore, we set $\kappa_1 =  \gamma/2$ without  this choice having any impact on our analysis. Furthermore, the value of $\kappa_2$ and $\gamma$ do not need to be specified in the calculation at all. Only $\Gamma$  needs to be specified. Third, we set $U=0.005$ $\Gamma$. This choice determines the number of photons $|\alpha|^2 $ involved in the bistability and the critical driving amplitude $F_c$ needed to reach the bistable regime. Note, however, that any spectral lineshape can be attained for any value of $U$ by scaling $F$ (which determines $|\alpha|^2 $) accordingly. Therefore, our choice of $U$ doest not impact our analysis. Based on the above considerations and the choice of $U$, we are left with $F$ as the only adjustable parameter used to fit the calculated lineshape to the measured lineshape. At this point, we would like to note an additional constraint related to the critical driving amplitude $F_c$ for which bistability emerges.  In particular, for $F=F_c$ we have $U |\alpha|^2 \sim \Gamma$. Experimentally, we can  estimate $F_c$ by performing dynamic hysteresis measurements at different powers. $F_c$ then corresponds to the minimum laser amplitude for which bistability is observed. Consequently, any laser amplitude can be referenced to $F_c$. In this way, we find that the value $F/F_{c}=1.52$ (as obtained from the calculations) giving the best fit to the experimental data is fully consistent with  our experimental estimate of $F/F_c \approx 1.5$. Finally, we note that the fact that $F/F_c$ is the relevant way to express the driving amplitude in our mean-field model, combined with the fact that $\kappa_1$ determines $F_c$, further supports the statement that the value of $\kappa_1$ we selected is irrelevant to our analysis of the spectral lineshape.

For dynamic hysteresis calculations based on Equations 2 and 3 in the main text, we use all of the same parameter values mentioned above. In addition, we set the thermal relaxation time to $\tau=10^{4}$ $\Gamma^{-1}$. Our choice of the value of $\tau$, which is somewhat shorter than the experimental one as explained in the main text, was based on two considerations. First, calculations using the experimental value of $\tau$ would be extremely long and memory-expensive. Second, as long as $\tau \gg \Gamma^{-1}$, the physics of dynamic hysteresis in our single-mode cavity remains qualitatively the same. A longer $\tau$ will simply shift the maximum hysteresis area to slower scans, but it will not change the shape of the hysteresis area curve as a function of the scanning speed. Therefore, we selected a sufficiently  large value of $\tau$ that still allows us to perform the calculations within a reasonable ($\sim$ days) time.

The scans we performed to calculate the hysteresis area consist of varying the detuning $\Delta$ in a linear and symmetric way from $-20$ $\Gamma$ to $20$ $\Gamma$.  Finally, since the equations of motion are deterministic, it is sufficient to calculate the response for only one period for each $T$. The calculation results are presented in Figures 3b, 4, and 5, of the main article.

\end{document}